\documentclass[OL,preprint,showpacs,superscriptaddress,12pt]{revtex4-1} 
\usepackage{amsmath,amssymb,graphicx}
\begin{document}

\title{Enhanced propagation of  photon density waves  in random amplifying media }

\author{Lalruatfela Renthlei}
\author{Harshawardhan Wanare}  \email{Corresponding author: hwanare@iitk.ac.in}
\author{S. A. Ramakrishna}
\affiliation{Department of Physics, Indian Institute of Technology, Kanpur 208016 INDIA}

\begin{abstract}
We demonstrate enhanced wave-like character of  diffuse photon density waves (DPDW) in an amplifying random medium.
The amplifying nature makes it contingent to choose the wave solution that grows inside the amplifying medium, and has a propagation
vector pointing opposite to the growth direction. This  results in negative refraction of the DPDW  at an absorbing-amplifying random medium interface as well as the possibility of supporting  ``anti"-surface-like modes at the interface.  A slab of amplifying random medium sandwiched between two absorbing random media supports waveguide resonances that can be utilized to extend the 
imaging capabilities of DPDW.
\end{abstract}

\maketitle 
It is well known that a time-modulated light source placed in a random scattering medium creates
a modulation in the density of the diffuse photon flux~\cite{yodh1}. These density variations propagate 
in the source free regions of the random medium according to the diffusion equation 
\begin{equation}
\vec{\nabla} \cdot (D \vec{\nabla} \phi) - \mu_a \phi - \frac{1}{c} \frac{\partial \phi}{\partial t} = 0,
\label{diffusion}
\end{equation}
where, $D= 1/3(\mu_s^\prime+\mu_a)$ is the diffusion coefficient, with $\mu_s^\prime$ being the reduced scattering coefficient, $\mu_a$ is the absorption 
coefficient,  $\phi$ is the amplitude of the density variation and $c$ is the speed of light in the medium in which the scatterers are embedded.  The waves have been 
popularly known as Diffuse Photon Density Waves (DPDW) in the area of biomedical imaging where it has been used for 
a variety of imaging through biological media~\cite{yodh, durduran}. 
These waves are known to exhibit  interference~\cite{inter} and 
diffraction~\cite{boasdiff}, and can be related across interfaces between two distinct scattering media by reflection and transmission coefficients~\cite{yodh1,ripoll}.
DPDW suffer from strong attenuation due to scattering and typically have penetration depths of only a couple of 
centimeters at $\sim100$~MHz modulation frequencies.

The diffusion approximation itself breaks down at much higher modulation frequencies \cite{harsha_su_grobe}. 
We note that other theories for photon migration \cite{weiss,sar} have been shown to be advantageous in the short time limits (large modulation frequencies). However, the diffusion equation \ref{diffusion} describes the DPDW well in the steady state at length scale
much larger than the transport mean free path for reasonable modulation frequencies ($ < 200$ MHz) \cite{yodh1,yodh}.
Another
 limitation for the DPDW intensity  arises from the intrinsic absorption in the medium. 
 Amplification of light  has been  used to compensate for attenuation
of the waves in many contexts, for example,  fiber amplifiers \cite{edfa-book}, amplifying media in 
metamaterials/perfect lenses of negative refractive index \cite{zeludev,sar1}, extending surface plasmon propagation
lengths at the interface of metal/amplifying medium \cite{grafstrom}. Random amplifying media are well known in the context of 
random lasers \cite{cao_iop}. The compensation of absorption by laser gain  in random media is known to  lead to the sharpening of the 
coherent backscattering peak and increased weak localization \cite{wiersma}.  

 \begin{figure}[htb]
 \centering
  \includegraphics[width=0.9\columnwidth]{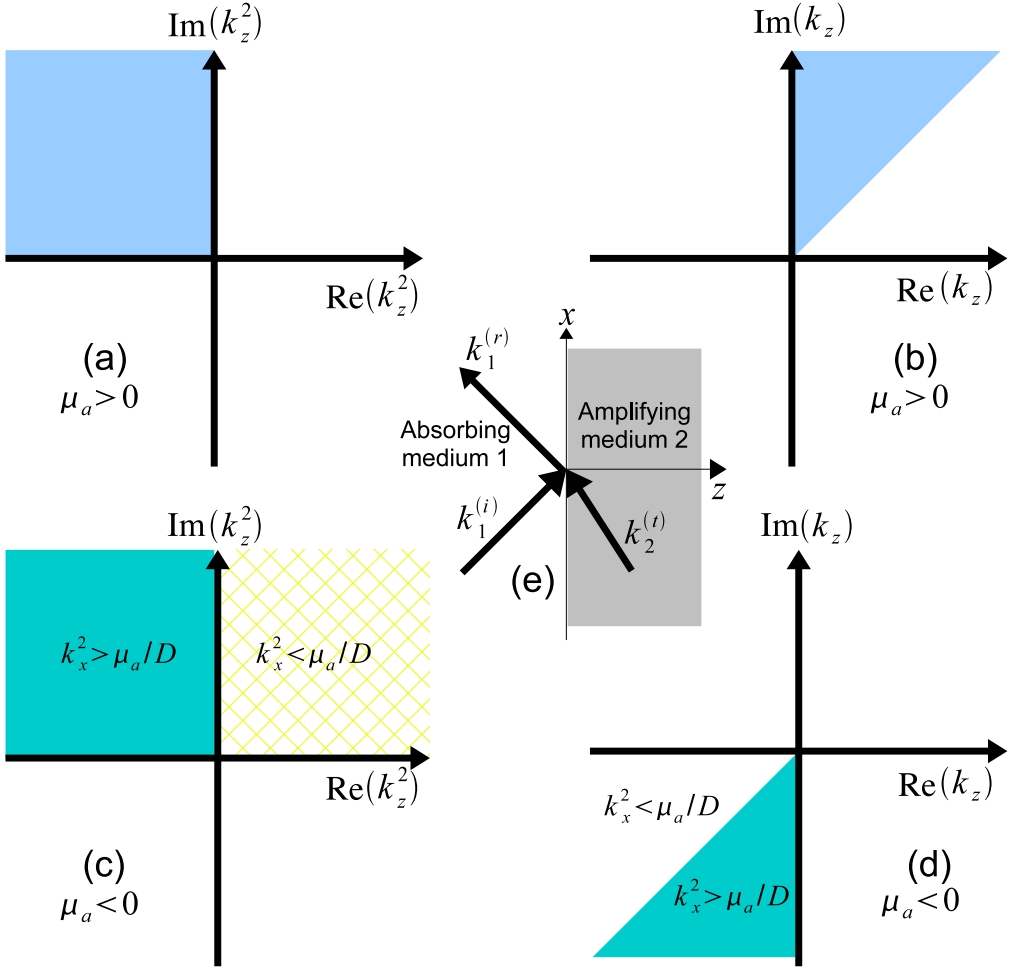}
 \caption{ \label{fig1} The highlighted regions indicate all possible values that the propagation vector ($k^2$ and $k$)  can
 take in the complex plane, (a,b) and (c,d) show these for  absorbing and amplifying random media, respectively. (e) Schematic of the incident,
 reflected and the transmitted wavevectors at the interface of absorbing-amplifying random media ($z=0$ plane).}
\end{figure}

 In this letter, we examine the behaviour of  DPDW in an amplifying random medium. 
 The incorporation of gain in the random medium results in extending the wavelike characteristics of the DPDW, wherein the
 real part of the wavevector  can become larger than the imaginary part.
  For an incident DPDW at an interface of
 absorbing-amplifying media, we find that negative refraction of the DPDW can occur. This results from a physical choice 
 of the wavevector that is mandated by the requirement of a growing density wave  inside the amplifying random medium
 in spite of the DPDW having a  positive phase velocity in both the media.
 Although the finite modulation frequency of the DPDW renders the waves exponentially decaying, a slab 
 of amplifying random medium can support resonant waveguide modes.

 Assuming plane wave solutions of the form, $\phi \sim \phi_0 \exp[i~(\vec{k} \cdot \vec{r} - \omega t)]$, for
 the diffusion equation results in the following dispersion relation
 \begin{eqnarray}
  k^2 &=& \frac{-c \mu_a +i\omega}{cD}. \label{ksqr}
  \label{dispersion}
 \end{eqnarray}
In order to obtain the wave-vector $k$, we are faced with the classic problem of choosing the sign of the square root 
\cite{sar_om_OL_2005}
from the dispersion relation in Eqn. \ref{dispersion}, which yields
\begin{equation}
k = \pm \left[ \frac{-c \mu_a +i\omega}{cD}\right]^{1/2}~. \label{root}
\end{equation}
Note that the  coefficient ($\mu_a$) can either be positive (for absorption) or negative (for amplification).  
The validity of the diffusion equation to describe photon migration is contingent on the magnitude $|\mu_a| \ll \mu_s^\prime$, 
whereby scattering dominates over absorption/amplification.
The level of amplification of the diffuse flux due to $\mu_a<0$ establishes the threshold for lasing in random lasers \cite{cao_iop} or criticality in nuclear reactors.
The imaginary part of the wave-vector arises principally due to the non-zero frequency of modulation.
The correct physical procedure to fix the sign lies in studying the wave in a \textit{dissipative}  
random medium and choosing a  wave-vector  that causes decay of the wave with increasing distance of 
propagation $[\mathrm{Im}(k) > 0]$.  For the time being, we focus our attention on the complex number 
$k$ in Eqn.(\ref{root}). We note that $k^2$ lies in the second quadrant of the complex plane and hence $k$ in the absorbing medium  
will lie in the first quadrant (see Fig. 1).
This gives rise to the usual  DPDW that exponentially decays with distance due to an imaginary 
part of $k$ that is larger than the real part of $k$, thereby limiting the wavelike character.

 \begin{figure}[htbp]
  \begin{minipage}[b]{0.8\columnwidth}
    \centering
    \includegraphics[width=\columnwidth]{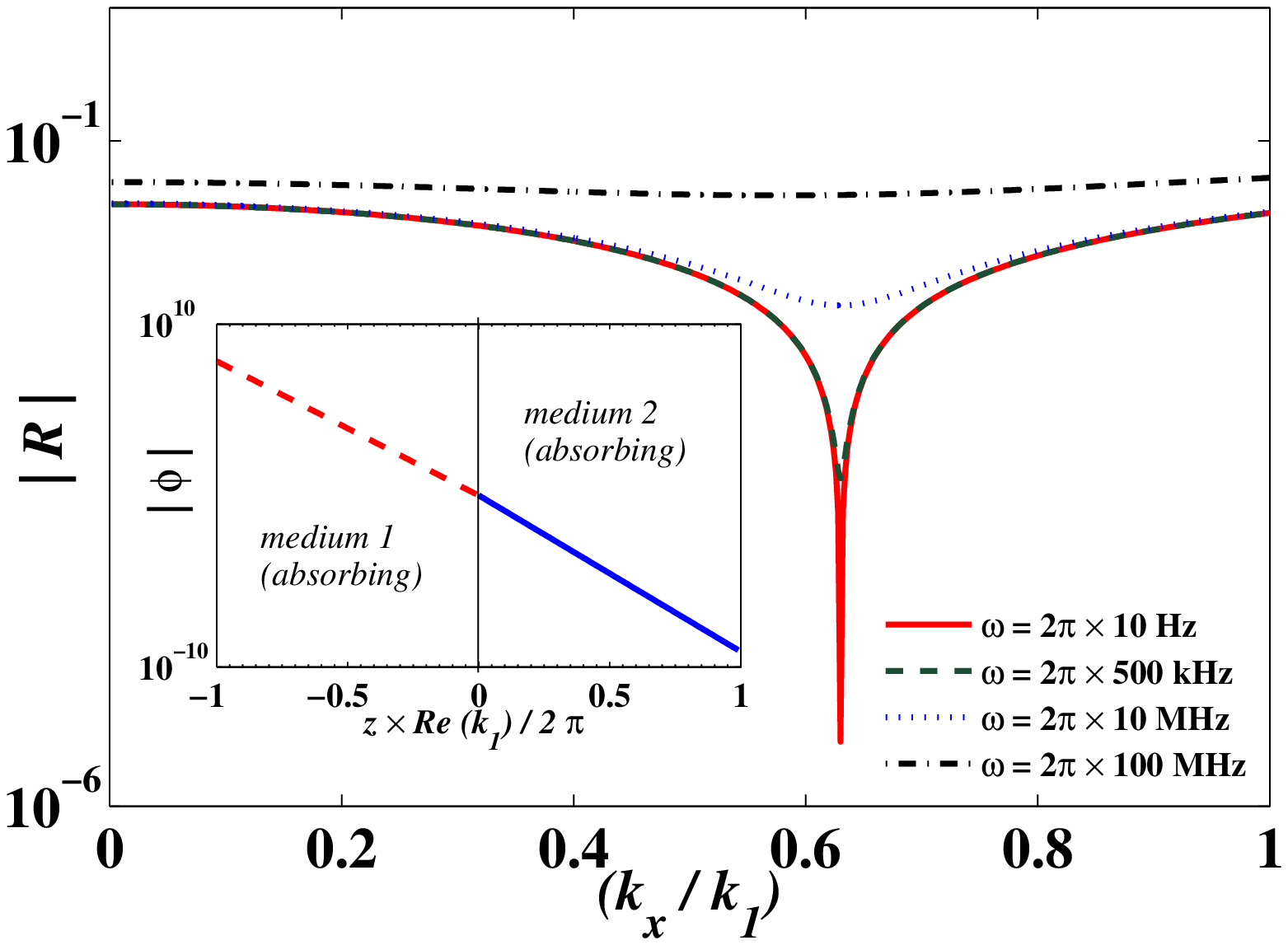}
  \end{minipage}
  \begin{minipage}[b]{0.8\columnwidth}
    \centering
    \includegraphics[width=\columnwidth]{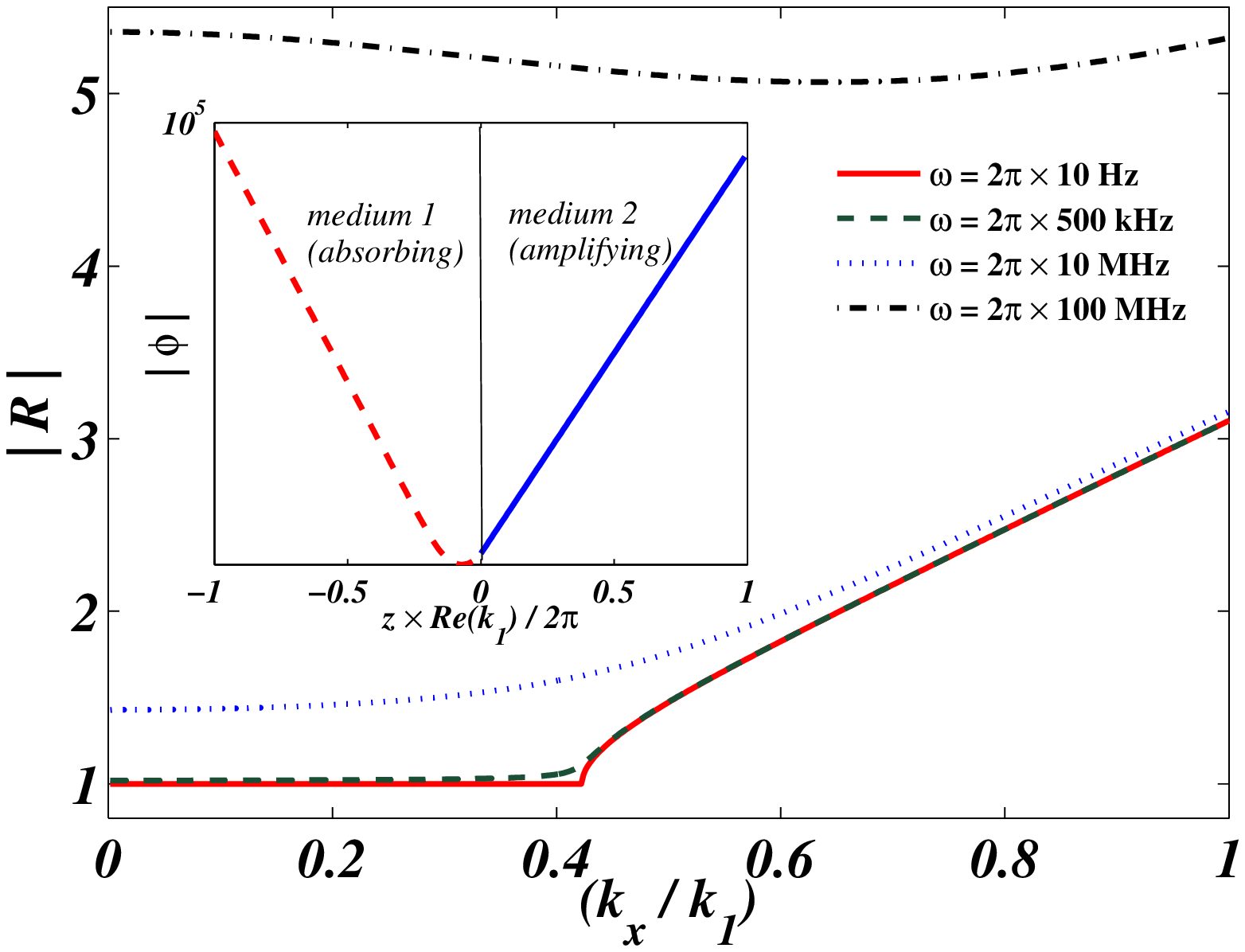}
  \end{minipage}
   \caption{  \label{fig2} (a) The reflection of a DPDW at an absorbing-absorbing random medium interface results in a Brewster-like condition, with $\mu_{a1} = 0.0175~\mathrm{cm}^{-1}$, $D_1=0.0416~\mathrm{cm}$, $\mu_{a2} =0.025~\mathrm{cm}^{-1} $, and $D_2=0.033~\mathrm{cm}$. (b) The reflection of a DPDW at an absorbing-amplifying random medium interface is shown for $\mu_{a2} =-0.0025~\mathrm{cm}^{-1} $, and $\mu_{a1}$, $D_1$ and $D_2$ are same as in (a). The insets indicate the DPDW amplitude across the interface in the two media at $\omega = 2\pi \times 100~{\mathrm  MHz}$.}
\end{figure}

We now turn our attention to an \textit{amplifying} random medium. 
In this case, $k^2$  lies in the first quadrant of the complex plane and $k$ could lie either in the first quadrant or in 
the third quadrant.
Due to amplification, the amplitude of the diffuse photon flux will increase exponentially 
with increasing distance of propagation under conditions of linear unsaturated gain. Hence, we require the wavevector to have a 
negative imaginary part  $\mathrm{Im}(k) <0$  for  a physical solution, and  $k$ will lie in the third 
quadrant as shown in Fig. \ref{fig1} ($k_x = 0$ for our discussion here). 
Note,  that regardless of the choice of the sign of $k$, the $\mathrm{Re}(k)$ 
is always greater than the $\mathrm{Im}(k)$ for the random amplifying 
medium. This implies that the oscillatory aspect of the DPDW is enhanced over the exponential decay arising intrinsically from multiple scattering, which also contributes to amplification owing to the resulting longer path through the gain medium.  A discussion of the relevant physical length and frequency scales is imperative at this juncture: the amplifying medium provides unbiased 
amplification of both the background diffuse flux as well as the flux associated with the DPDW, as these two 
fluxes cannot be distinguished by the stimulated emission process. The modulation frequency (typically $10$ to $100$ MHz) responsible 
for the DPDW is much smaller compared to the bandwidth of the gain medium ($\sim $ THz for typical dye molecules).
The gain can be considered uniform as the diffusion length scales are much shorter than the wavelength of the DPDW.

Now let us understand the solutions with our requirement for $\mathrm{Im}(k)<0$ for the random amplifying medium which involves 
 a negative real part of the wave-vector. 
The oscillatory flux associated with the DPDW given by $\vec{J} = -D \vec{\nabla} \phi$, points in the same direction as the real part of the 
wavevector.  
The complex nature of the the $\vec J$ arises due to the complex phasor notation for the PDW used in the paper. The real and imaginary parts should simply be viewed as representing the in-phase and in-quadrature components of the  response in relation to the oscillatory source term. 
We have a wave with positive phase velocity where both the wavevector and the flux vector point in the 
same direction. Both these consequences are dictated
by the need for having the DPDW amplitude  as well as the background diffuse flux that exponentially increase with distance from 
a source in a gain medium.
This solution should be compared with solutions pertaining to the electromagnetic waves  in the negative refractive index media 
\cite{sar_book}. A negative choice of the wavevector (real part) is required in that context also. However, the Poynting vector
representing the energy flux associated with the electromagnetic wave is oppositely oriented to the  wavevector, which is
an intrinsic  character of a wave with negative phase velocity. Note that these differences principally arise from the governing
equations: the Maxwell's equations for the electromagnetic waves and the diffusion equation for the DPDW.

 \begin{figure}[htbp]
 \begin{minipage}[thb]{0.7\columnwidth}
    \centering
    \includegraphics[width=0.8\columnwidth]{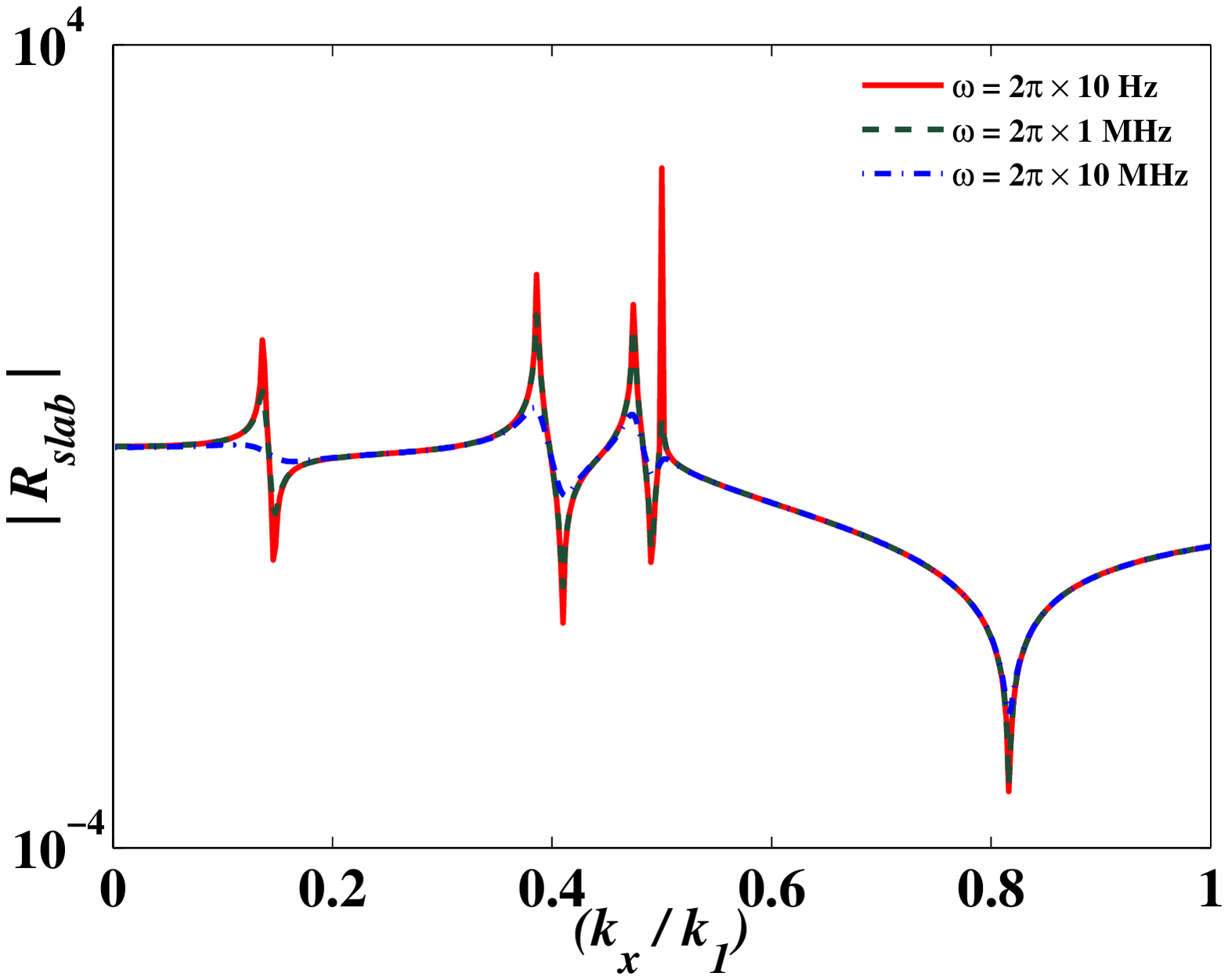}
  \end{minipage}
  \begin{minipage}[b]{0.7\columnwidth}
    \centering
    \includegraphics[width=0.8\columnwidth]{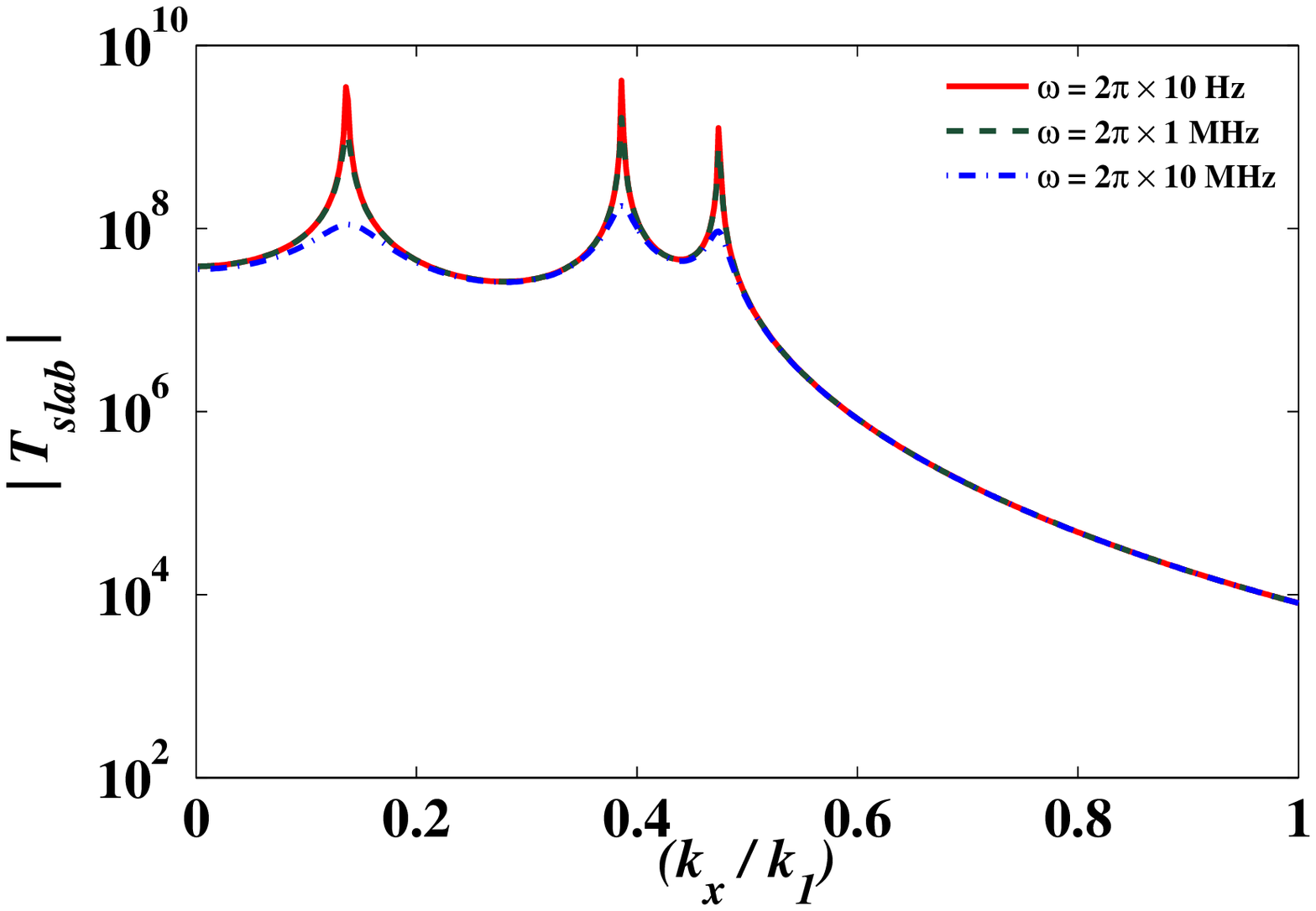}
  \end{minipage} 
   \begin{minipage}[b]{0.7\columnwidth}
   \centering
    \includegraphics[width=0.8\columnwidth]{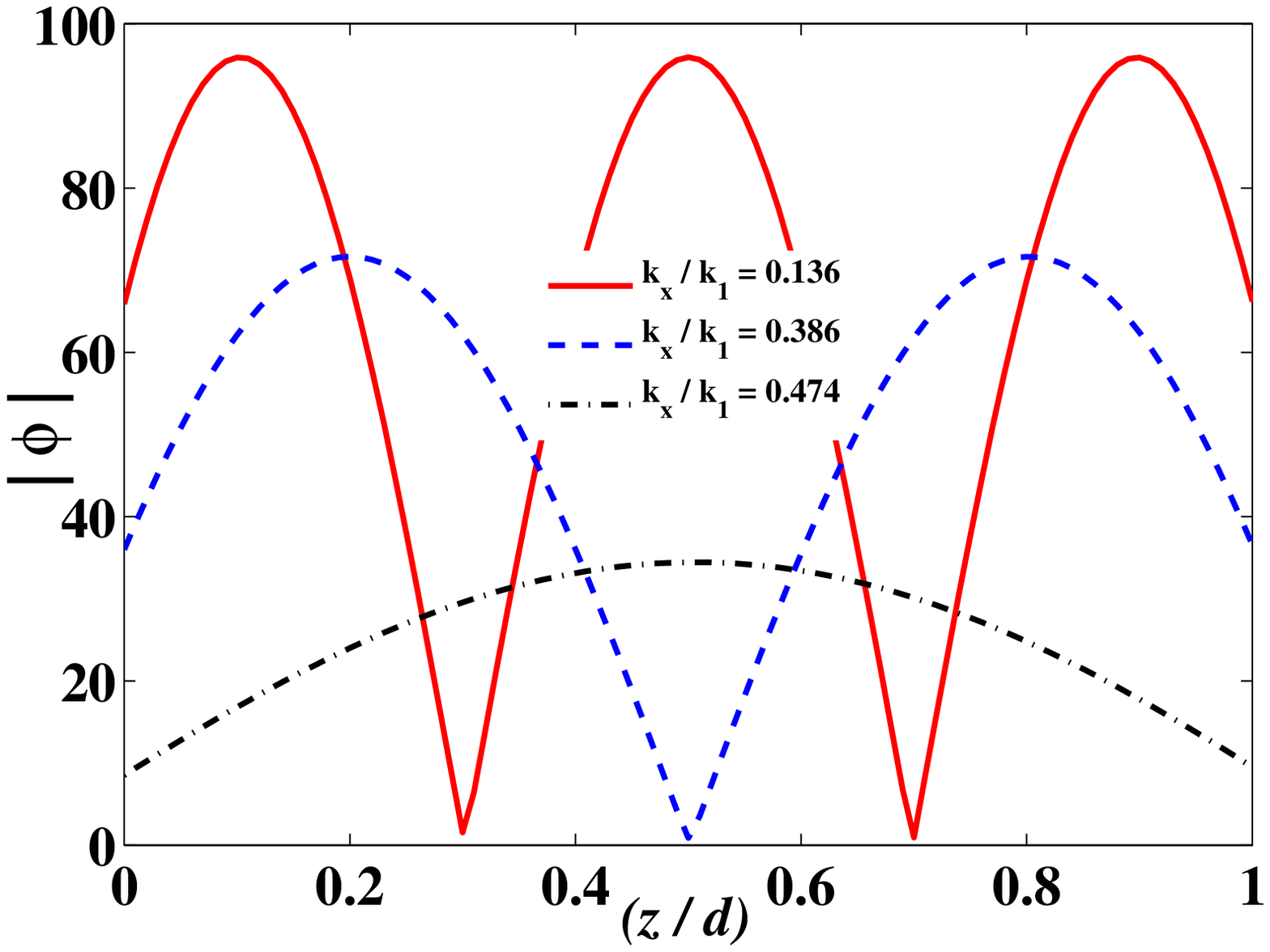}
  \end{minipage}
   \caption{  \label{fig3} For a slab of amplifying random medium with $\mu_a = -0.1~\mathrm{cm}^{-1}$ and $D=0.33~\mathrm{cm}$ sandwiched between two identical absorbing random media with  $\mu_a = 0.2~\mathrm{cm}^{-1}$ and $D=0.166~\mathrm{cm}$, the reflection and transmission are shown in (a) and (b), respectively. (c) The
 amplitude of the DPDW within the slab at the transmission peaks indicated in (b) for $\omega = 2\pi~{\mathrm MHz}$.}
\end{figure}

Consider the refraction of the DPDW at an interface between an absorbing random medium and an amplifying random 
medium when the DPDW  is incident from the absorbing side.
Without  loss of generality, we choose the $z = 0$ plane to be the interface and invariance along the $y-$direction, so that the 
incident wavevector lies in the $x-z$ plane such that  
$  k_x^2 + k_z^2 = k^2 = {(-c \mu_a +i\omega)}/{cD}$ (see Fig. \ref{fig1}e). 
Continuity of the interface in the $x-$direction implies equality of 
$k_x$ in both the media. The normal component $k_z$ is given by 
\begin{equation}
k_z = \pm \left[ \frac{-c( \mu_a + D k_x^2) +i~\omega}{cD}\right]^{1/2}, 
 \end{equation}
with the choice of sign in the two media governed by the previous discussion. 
The evidence of negative refraction at the interface lies in the decrease of the phase of the DPDW with increasing distance from the interface into the amplifying medium.
Ensuring the conditions of continuity of the 
amplitude of the DPDW and the normal component of the flux yields
\begin{eqnarray}
\phi_{\small I} + \phi_{\small R} = \phi_{\small T} & \rightarrow& 1+R = T\nonumber \\
(\vec{J_{\small I}} + \vec{J_{\small R}} ) \cdot \hat{z} = \vec{J_{\small T}} \cdot \hat{z} &\rightarrow&   D_1 k_{z1} (1-R) = D_2 k_{z2} T
\end{eqnarray}
 where the subscripts {\small {I,R}} and {\small T} refer to the incident, reflected and the transmitted DPDW at the interface, respectively.
The reflection and transmission coefficients are given by 
\begin{equation}
 R  =  \frac{\phi_{\small R}}{\phi_{\small I}} = \frac{D_1 k_{z1} - D_2 k_{z2}} {D_1 k_{z1} + D_2 k_{z2}},  
~~T  
 =\frac{2 D_1 k_{z1}} {D_1 k_{z1} + D_2 k_{z2}}.
\label{reflect}
\end{equation}

Ripoll and Nieto-Vesperinas  \cite{ripoll} have shown the existence of a Brewster-like condition, where the reflection coefficient ($R$) 
goes to zero, for a DPDW with $\omega \rightarrow 0$ (the static limit). This condition can be 
thought of as arising from matching of the absorption 
length in the two media with an appropriately factored periodicity of the photon density wave along the propagation direction $z$.
This {\em impedance} matching occurs at $k_x$ at which $D_1 k_{z1} = D_2 k_{z2}$, see Fig. \ref{fig2}(a), which in turn is determined by the various parameters of the medium, where 
\begin{equation}
k_x^2 = \frac{ \mu_{a1} D_1 -  \mu_{a2}D_2}{D_2^2-D_1^2}  +\frac{i\omega}{D_1 + D_2}. 
\label{kxx}
 \end{equation}
For $\omega \ll {\mu_a/D}$ the imaginary part is negligible and 
a proper choice of $\mu_a$ and $D$ allows one to tailor its occurrence and this can be used to realize a directional filter~\cite{ripoll}.

It appears intriguing that a pair of rather isotropic random media should choose a specific $k_x$ for an efficient coupling of the 
DPDW.  The conventional Brewster angle observed for  electromagnetic waves reflected across interfaces is a polarization dependent feature that arises from the requirement of orthogonality of the   induced material polarization to the reflected direction. 
 What would then underly the Brewster-like condition of the  scalar density waves?
The above Eqn.(\ref{kxx}) indicates that it is the matching of the absorption length  $l_{\mathrm{abs}} = \sqrt{l_s l_a/3}$ across the two media that  is more fundamental, and hence the difference between the $l_{\mathrm{abs}}$ at the interface of these media 
is augmented by the 
$k_x$ component of the DPDW. Here, $l_s = \mu_s^{\prime^{-1}}$, $l_a = \mu_a^{-1}$ and the absorption length $l_{abs} $ is the (rms)
average distance between begin and end points of paths of length $l_a$ in the random medium. Thus, the reflection minimum leading to strong coupling occurs such that the wave propagation  dictated by the periodicity of the $k_x^{-1}$  is matched with the  difference of the absorption lengths in the two media.

In the case of the  absorbing-amplifying media interface, the above situation simply does not exist. 
The sign of ${k_{z2}}$ changes, and the expressions of $R$ and $T$ Eqn.(\ref{reflect})
transform appropriately, leaving behind a denominator that is positive definite because of the reversal of the sign of 
$\mu_{a2}$ of the amplifying medium \cite{foot}. For  $\omega \ll {|\mu_a|/D}$ one obtains  $R \sim 1$ at normal incidence
as well as for  $k_x^2< {|\mu_a|/D}$. At $k_x^2 = {|\mu_{a2}|/D_2}$, the real part of $k_{z2}$ becomes zero 
and beyond this point the  reflection coefficient continues to increase  as the input angle is increased
due to the dominance of the imaginary part of the wavevector (Fig. \ref{fig1}d).
Note again that the pertinent length scale is the amplification length ($l_{\mathrm{amp}}$ defined similarly to the 
$l_{\mathrm{abs}}$ above) which determines the condition on the different regimes of $k_x$.
In Fig. \ref{fig2}(b) we present reflection and transmission coefficients at an interface of  an absorbing-amplifying medium for various modulation frequencies.

One can also arrange to obtain an \textit{``anti"-surface} mode at an absorbing-amplifying interface having a minima at
the interface and the field growing with distance on both sides of the interface (see inset in Fig.\ref{fig2}b). 
These fields do not correspond to a true surface mode as the DPDW diverges at infinity, and hence cannot exist in isolation
without the exciting input density wave.
Such a mode has been envisaged in the context of impedance matched perfect lenses made of a negative refractive index 
material \cite{pendry-sar}. This physical realization can occur only in conjunction with an  amplifying medium and 
an {absorbing-absorbing} interface cannot support such a mode. 

We present the response of a slab of amplifying random medium sandwiched between identical 
absorbing random media, with the reflection and transmission given as 
\begin{eqnarray}
R_{\mathrm{slab}} &= & \frac{2i}{\Delta} [(D_1k_{z1})^2 - (D_2 k_{z2})^2] \sin{k_{z2}d}, \\
T_{\mathrm{slab}}&=& -\frac{4}{\Delta}D_1D_2 k_{z2}  k_{z1} e^{-i k_{z1}d}  ,
\end{eqnarray}
where,  $d$ is  the thickness of the slab and $\Delta = (D_1 k_{z1} - D_2 k_{z2})^2 e^{ik_{z2}d}-(D_1 k_{z1} + D_2 k_{z2})^2 e^{-ik_{z2}d}$. The above expressions for a finite slab are invariant with respect to the sign of $k_z$
\cite{axl4}.
The reflection and transmission as a function of $k_x$ show large peaks at the resonant conditions for an amplifying slab
as in Fig. \ref{fig3}.
Such a configuration leads to \textit{localized} waveguide modes within the amplifying slab. 
The density variation
with distance inside the slab is shown in Fig. \ref{fig3}(c). The integral half-wavelength-like conditions are apparent. 
We have chosen the parameters in Fig.\ref{fig3} such that $|k_1|>|k_2|$ and hence for $k_x>|k_2|$ the field
in the slab is evanescent. Due to this, the transmission does not contain resonant features for $k_x \ge |k_2|$.
The ability of  a layer of  amplifying  random medium to guide DPDW  can be exploited in imaging situations, wherein coupling to the random medium and  extraction of the DPDW  can be greatly enhanced.

In conclusion, we have demonstrated that incorporating amplification (laser gain) in a random medium can substantially enhance
the wavelike characteristics of the DPDW. This aspect may be used for increasing the depth over which imaging can be carried out in a random medium with DPDW. The physical requirement of a growing density wave in an amplifying random medium  results in a backward
flux towards the interface with the absorptive random medium. This gives rise to negative refraction across the interface and 
conditions can result to produce ``anti"-surface like states at such interfaces. A slab of amplifying random medium is shown 
to waveguide DPDW and we suggest that such waveguides can be used  for  improving the imaging capabilities of the DPDW, whose non-invasive aspects continue to hold great promise in diagnostic imaging.

We thank S. Guenneau for initial discussions on possibility of negative refraction in diffusive systems.

\end{document}